\documentclass[aps,superscriptaddress,prl,reprint, linenumbers, showpacs
,amsmath,amssymb,twocolumn]{revtex4-1}

\usepackage{graphicx}
\usepackage{dcolumn}%
\usepackage{bm}
\usepackage{color} \usepackage{amsmath} \usepackage{lineno}
\usepackage{url}

\newcommand{\CV}{\operatorname{\mathsf{CV}}}

\begin{document}

\title{Adaptation Reduces Variability of the Neuronal Population Code}


\author{Farzad Farkhooi}
\email[Corresponding author: ]{farzad@zedat.fu-berlin.de}
\affiliation{Neuroinformatics \& Theoretical Neuroscience, Freie Universit{\"a}t
Berlin and BCCN-Berlin, Germany}

\author{Eilif Muller}
\affiliation{Laboratory of Computational Neuroscience, EPFL, Lausanne,
Switzerland}

\author{Martin P. Nawrot}
\affiliation{Neuroinformatics \& Theoretical Neuroscience, Freie Universit{\"a}t
Berlin and BCCN-Berlin, Germany}

\date{\today}

\begin{abstract}
  Sequences of events in noise-driven excitable systems with slow
  variables often show serial correlations among their intervals of
  events.
Here, we employ a master equation for generalized non-renewal processes to
calculate the interval and count statistics of superimposed processes
governed by a slow adaptation variable.
%
For an ensemble of spike-frequency adapting neurons, this results in
the regularization of the population activity and an enhanced
post-synaptic signal decoding. We confirm our theoretical results in a
population of cortical neurons recorded \emph{in vivo}.
%
\end{abstract}



\maketitle

Statistical models of events assuming the renewal property, that the
instantaneous probability for the occurrence of an event depends
uniquely on the time since the last event, enjoys a long history of
interest and applications in physics.
%
However, many event processes in nature violate the renewal
property. For instance, it is known that photon emission in multilevel
quantum systems constitutes a non-renewal process \cite{*[{See
      Refs. [6-7] and [19-26] in }] soler_memory_2008}. Likewise, the
time series of earthquakes typically exhibits a memory of previous
shocks \cite{livina_memory_2005}, as do the times of activated escape
from a metastable state, as encountered in various scientific fields
such as chemical, biological, and solid state
physics~\cite{lindner_correlations_2007}.
%
%
Often, the departure from the renewal property arises when the process
under study is modulated by some slow variable, which results in
serial correlations among the intervals between successive events.  In
particular, the majority of spiking neurons in the nervous systems of
different species show a serial dependence between inter-event
intervals (ISI) due to the fact that their spiking activity is modulated by
an intrinsic slow variable of self-inhibition, a phenomenon known as
spike-frequency adaptation
\cite{liu_spike-frequency_2001,*farkhooi_serial_2009}.

In this letter, we present a non-renewal formalism 
based on a population density treatment
%
%
that enables us to quantitatively study ensemble processes augmented
with a slow noise variable.
%
%
We formally derive general expressions for the higher-order interval
and count statistics of single and superimposed non-renewal processes
for arbitrary observation times. 
%
%
In spiking neurons, intrinsic mechanisms of adaptation reduce output
variability and facilitate population coding in neural ensembles. We
confirm our theoretical results in
%
%
a set of experimental \emph{in vivo} recordings and analyse their
implications for the read-out properties of a postsynaptic neural
decoder.

%
%
%

\textit{Non-renewal Master Equation.} We define the limiting
probability density for an event given the state variable $x$ by the
so-called hazard function $h_x(x, t)$ where $t$ denotes explicit dependence
 on time due to external input following
\cite{[{Supplementary material at}] supp,muller_spike-frequency_2007}.
%
%
Here, we assume $x$ has a shot-noise-like dynamics,
%
%
which is widely used as a 
model of spike induced
neuronal adaptation \cite{muller_spike-frequency_2007}
\begin{equation}
\textstyle \dot{x}: = - x(t)/\tau + q \sum_k \delta(t - t_k),
\label{SFAdynamics}\end{equation}
where $\delta$ is the Dirac delta function, $t_k$ is the time of the
$k^{\mathrm{th}}$ event, and $q$ is the quantile change in $x$ at each
event.  The dynamics of $x$ deviates from standard treatments of
shot-noise (such as in \cite{gardiner_book}) in that the rate of
events has a dependence on $x$ as expressed by the hazard function
$h_x(x,t)$.  It is straightforward to show that the distribution of $x$ in
an ensemble, denoted by $\Pr(x,t)$,
is governed by 
%
\begin{eqnarray}
 \partial_t \Pr(x, t) &= & \partial_{x} [\frac{x}{\tau} \Pr(x, t) ] +
 h_x( x - q, t)\Pr(x - q, t) \nonumber \\ & - & h_x( x , t)\Pr(x , t).
\label{conME1}\end{eqnarray}
%
%
%
Much insight can be gained by applying the method of characteristics
\cite{dekamps_2003} to establish a link between the state variable $x$
and its time-like variable $t_x$. 
For Eq.~\eqref{SFAdynamics} 
we define $t_x = \eta (x) := -\tau \ln(x/q)$, whereby $\frac{d}{dt}t_x
= 1$.  When an event occurs, $t_x \mapsto \psi(t_x)$, where $\psi(t_x)
= \eta( \eta^{-1}(t_x) + q) = - \tau \ln ( e^{ -t_x/\tau} + 1)$ with
its inverse given by $\psi(t_x)^{-1} = - \tau \ln ( e^{ -t_x/\tau} -
1)$.
Thus, we define $h(t_x, t) := h_x(\eta^{-1}(t_x), t)$.
This transformation of variables to $t_x$ elucidates the connection of
the model to renewal theory.
%
Here, the reset condition after each
event is not $t_x \mapsto 0$ (renewal) but $t_x \mapsto \eta( x + q)$
\cite{[{Supplementary material at}] supp}. Therefore, the variable
$t_x$ that we may call a 'pseudo age' is a general state variable that
no longer represents the time since the last event (age).
%
Transforming variables in Eq.~\eqref{conME1} from $x$ to $t_x$
yields in the steady state
%
%
\begin{eqnarray}
 \partial_{t_x} \Pr(t_x) &=&- h(t_x)\Pr(t_x) \nonumber \\ &+& (1 -
 \Theta_0(t_x)) [h(\psi^{-1}(t_x) \Pr(\psi^{-1}(t_x)) ],
\label{staticME1re}\end{eqnarray}
where $\Theta_0(t_x)$ is the Heaviside step function, and for
convenience we defined $\psi^{-1}(t_x \geq 0 ) \equiv 0$.
An efficient algorithm for solving Eq.~\eqref{staticME1re} is given in
\cite{muller_spike-frequency_2007}.
%
We denote this solution by ${\Pr}_{eq}(t_x)$.
Further, the time-like transformation in Eq. \eqref{staticME1re}
allows computation of the ISI by analogy to the renewal
theory \cite{muller_spike-frequency_2007} and also permits the
comparison to the master equation for a renewal process as given
in Eq. (6.43) in \cite{gerstner_spiking_2002}.
%
The distribution of $t_x$ just prior to an event
is a quantity of interest and 
it is derived as ${\Pr}^*(t_x) = h(t_x){\Pr}_{eq}(t_x)/r_{eq}$, where
$r_{eq} = \int h(t_x) {\Pr}_{eq}(t_x) dt_x $ is a normalizing constant
and also the process intensity or 
rate of the ensemble.
%
%
%
%
%
%
%
Similarly, one can derive the distribution of $t_x$ just after the
event, $\Pr^{\dagger}(t_x) = \Pr^*(\psi^{-1}(t_x)) \frac{d}{d
  t_x}\psi^{-1}(t_x)$ \cite{muller_spike-frequency_2007}.
Then the relationship between $t_x$ and the ordinary ISI
 distribution can be written as
%
\begin{equation}
 \rho(\Delta) = \int_{-\infty}^{+\infty} h(t_x+ \Delta) \Omega( t_x +\Delta)
     {\Pr}^{\dagger}(t_x) dt_x,
\label{ISI}\end{equation}
%
%
where $\Omega( t_x +\Delta) = e^{-\int_{t_x}^{\Delta} h(t_x + u) du} $.
Now the $n^\mathrm{th}$ moment $\mu_n$ of the distribution and its
coefficient of variation $C_v$ can be numerically determined.
%

\textit{Counting Statistics}.
%
%
%
In order to derive the count distribution,
%
we generalize the elegant approach for deriving the moment generating
function as introduced in \cite{van_vreeswijk_stochastic_2010}: let
$\rho_n ( t_n, t_x^n | t_x^0)$ be the joint probability density given
its initial state $t_x^0$, where $t_n$ stands for time to
$n^\mathrm{th}$ event and $t_x^n$ is the corresponding adaptive state
of the process. Thereafter, one can recursively derive
\begin{equation}
\tilde{\rho}_{n+1}(s , t_x^{n+1} | t_x^0)=  \int \tilde{\rho}_{n} (s, t_x^n
| t_x^0) \tilde{\rho}(s , t_x^{n+1} | t_x^{n}) dt_x^{n},
\end{equation}
where $\tilde{\rho}_{n+1}(s , t_x^{n+1} | t_x^0) = \mathcal{L}
[\rho_{n+1}(t_{n+1} , t_x^{n+1} | t_x^0)]$ and $\mathcal{L}$ is the
Laplace transform with resepect to time, assuming
$\tilde{\rho}_{1}(s , t_x^{1} | t_x^0)= \tilde{\rho}(s , t_x^{1}
|t_x^0)$ \cite{van_vreeswijk_stochastic_2010}.
%
%
%
%
%
%
%
%
%
%
%
%
%
Next, defining the operator $\mathbf{P}_n(s)$ and applying Bra-Kat
notation as suggested in~\cite{van_vreeswijk_stochastic_2010}, leads to the 
Laplace transform of $n^\mathrm{th}$ events ordinary density
%
%
\begin{equation}
\textstyle \tilde{\rho}_{n}(s) = \langle 1 \mid \mathbf{P}_n(s) \mid
           {\Pr}^{\dagger} \rangle = \langle 1 \mid [\mathbf{P}(s)]^n \mid
           {\Pr}^{\dagger} \rangle,
\label{ingenious2}\end{equation}
where the operator $\mathbf{P}$ associated with $\tilde{\rho}(s)$, which
interestingly corresponds to the moment generating function of the sum
of $n$ non-independent intervals $\tilde{f}_{n}(s)$ as defined
in~\cite{mcfadden_lengths_1962}. Now, following Eqs.~(2.15)
in~\cite{mcfadden_lengths_1962} Laplace transform of count
distribution denoted as  $\tilde{P}(n,s)$.


%
%



The Fano factor
%
provides an index for the quantification of the count variability. It
is defined as $J_{T} =\sigma_{T}^2/\mu_T$, where $\sigma_{T}^2$ and
$\mu_T$ are the variance and the mean of the number of events in a
certain time window $T$. It follows from the additive property of the
expectation that $\mu_T = \int_0^T r(u) du$ and assuming constant
firing rate $\mu_T = r_{eq} T $.
%
%
%
%
To calculate the second moment of $\tilde{P}(n, s)$, we require
$\tilde{\mathcal{A}}_s =  \sum_k \tilde{\rho}_{k}(s)$, thus
%
%
\begin{equation} 
\textstyle \tilde{\mathcal{A}}_s =
\langle 1 \mid \mathbf{P}(s) (\mathbf{I} -\mathbf{P}(s))^{-1} \mid {\Pr}
^{\dagger} \rangle,
\label{auto}\end{equation}
where $\mathbf{I}$ is the identity operator.
Note, assuming a renewal interval distribution in
Eq.~\eqref{ISI} one obtains $\tilde{ \mathcal{A}}^{r}_s = \tilde{\rho}(s) /
(1- \tilde{\rho}(s))$ and 
$\mathcal{L}^{-1}[r_{eq}\tilde{\mathcal{A}}_s] = r_{eq}\mathcal{A}(u)$ is
the joint density of an event at time $t$ and another event at time
$t+u$. Thus, the autocorrlation of events is $A(u) =
r_{eq}[\delta(u)+ \mathcal{A}(u)]$.
Now, by using Eq.~\eqref{auto} and the
Eq. (3.3) in~\cite{mcfadden_lengths_1962}, the second moment of the
count statistics can be derived.
%
%
%
Thus, we obtain the Fano factor
\begin{equation}
 \textstyle J_T = 1 + (2/T) \int_0^T (T - u) \mathcal{A}(u) du -
 r_{eq} T,
\label{Fano}\end{equation}
%
%
The asymptotic property of $F=\lim_{T \rightarrow \infty} J_T$ can
be derived from the result stated in Eq. (7.8) in
\cite{mcfadden_lengths_1962} as
\begin{equation}
 \textstyle \lim_{s\rightarrow 0} [ \tilde{ \mathcal{A}}_s - 1/(\mu_1
   s)] = C_v^2 [1/2 + \sum_{k=1}^{\infty}\xi_k] - 1/2,
\label{golden_eq}\end{equation}
where $\xi_k$ is the linear correlation coefficient between two $k$
lagged intervals.  Provided the limit exits, we find $F = C_v^2 [1 +
  2\sum_{k=1}^{\infty}\xi_k]$ in \cite{cox_statistical_1966}.
%

\textit{Superposition.} We now generalize our results on the counting
statistics to the superposition of independent point processes. This
is of practical interest in all cases where we observe 
superimposed events that stem from multiple independent process,
e.g.\ in photon detection devices, or in the case of a postsynaptic
neuron that receives converging inputs from multiple 
lines.
We study the superposition of $k$ stationary, orderly,
and independent processes. The ensemble process will have a 
%
%
%
rate $\check{r} = \sum_{i=1}^{k} r_i$ and following Eq. (4.18) in
\cite{cox_point_1980} $\check{\mathcal{A}}(u) = \check{r} +
\check{r}^{-1} \sum_{i=1}^{k} r_i[\mathcal{A}_i(u) - r_i]$. 
Here, for the sake of simplicity, we derive the desired relationship
between $C_v^2$ and the ensemble $\check{F}$ for $k$ identical
processes. To this end, we plug $\check{r}$ and
$\mathcal{L}[\check{\mathcal{A}}(u)]$ into the Eq. \eqref{golden_eq}
and therefore it becomes $\lim_{s\rightarrow 0} [
  \tilde{\mathcal{A}}_s - 1/(\mu_1 s)] = \CV^2[1/2 + \Xi] - 1/2$,
where $\CV$ and $\Xi =\sum_{i=1}^{\infty}\Xi_i $ are the coefficient
of variation and the interval correlations of the superimposed
process. Note that the left hand side of this equation and
Eq. \eqref{golden_eq} are simular. Thus, we obtian
\begin{equation}
\textstyle \CV^2[1 + 2\,\Xi] = C_v^2 [1 + 2\sum_{i=1}^{\infty}\xi_i].
\label{my_golden}\end{equation}
%
%
%
%
%
%
%
The left hand side of Eq.~\eqref{my_golden} is indeed the Fano factor
$\check{F}$ of the ensemble process as desired.  Now,
\cite{cox_point_1980} suggests as $k \rightarrow \infty$, $\CV^2
\rightarrow 1$. Interestingly, if all individual processes fullfill
the renewal condition, it follows from Eq.~\eqref{my_golden} that
$\check{F}=C_v^2 = [1 + 2\Xi]$, and therefore if $C_v^2 \neq 1$ the
population activity is non-renewal with $\Xi < 0$ ($\Xi > 0$) for
processes with $C_v^2>1$ ($C_v^2<1$). This important finding explains
the numerical observation in
\cite{lawrence_dependency_1973,*cteau_relation_2006,*lindner_superposition_2006}
regarding emergance of non-renewal processes as the result of the
superposition operation.

%
%
%
%

\textit{Adaptation in a Neuronal Ensemble.} 
In \cite{muller_spike-frequency_2007} it has been shown by an
adiabatic elimination of fast variables that the master equation
description of a detailed neuron model including voltage dynamics,
conductance-based synapses, and spike-induced adaptation reduces to a
stochastic point process simular to Eq. \eqref{staticME1re}.
%
 The corresponding hazard function can be approximated as
\begin{equation}
\textstyle \hat{h}_x(x) = a_t \exp(-b_t x),
\label{adap_h}\end{equation}
where $a_t$ and $b_t$ are determined by the time dependent statistics
of inputs \cite{[{Supplementary material at}] supp} and the
equilibrium rate consistency equation $ r_{eq} \approx
\hat{h}_x(r_{eq}q\tau)$~\cite{muller_spike-frequency_2007} with the
solution
\begin{equation}
\textstyle r_{eq} =\mathcal{W}(a b q \tau)/(b q \tau),
\label{r_con}\end{equation}
where $\mathcal{W}$ is the Lambert function.
%
%
%
In the case of vanishing adaptation ($b q \rightarrow 0$) the process
apporaches the Poisson process with $r_{eq}\rightarrow a$.


%

We show in \cite{[{Supplementary material at}] supp} that the
adaptation dynamics in Eq.~\eqref{SFAdynamics} produces negative
serial correlations $\xi_k < 0$.
The strength of serial correlation decays with increasing
lag $k$ and depends on the mean adaptation, $E[x] = r_{eq}q\tau$.
%
%
%
%
Such a vanishing of  
negative serial interval correlations with increasing lag is well
supported by a large body of experimental
evidence~\cite{farkhooi_serial_2009}.
The departure from the renewal property induced by adaptation reduces
the Fano factor Eq.~\eqref{Fano} for the single process as well as for
the population model of superimposed pocesses. 

%
%
%
%
%
%
\begin{figure}[t!]
\includegraphics[scale=1.0]{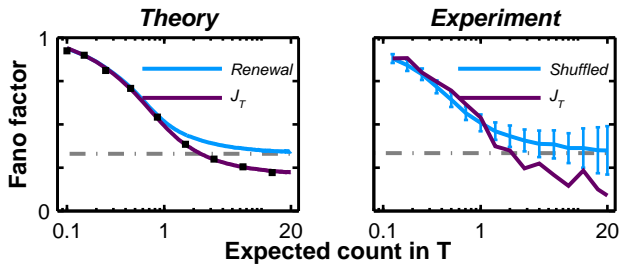}
\caption{Adaptation reduces the Fano factor of the ensemble process.
  \textbf{Left} {\it Magenta}: $J_T$ for arbitrary observation time
  $T$ according to Eqs.~\eqref{Fano} and~\eqref{adap_h} with $bq
  =1.4$, $a=5.0$ and $\tau =400ms$. {\it Blue}: Fano factor for
  equivalent renewal ensemble process with interval distribution of
  Eq.~\eqref{ISI}. {\it Square Dots}: Numerically estimated Fano factor
  for superposition of the 5 realization runs of the full-detailed
  adaptive neuron model as in \cite{muller_spike-frequency_2007}. {\it
    Dash-dotted line}: $C_v^2$.  \textbf{Right} {\it Magenta}:
  Empirical $\hat{J}_T$ estimated from the pooled spike trains of 5
  cortical neurons.  {\it Blue}: Fano factor for the pool of shuffled
  spike trains.  {\it Dash-dotted}: Average $C_v^2$ of the 5
  individual spike trains.}
\label{result}
\end{figure}

We validate our theoretical result of the reduced Fano factor
in a set of experimental spike trains of $N=5$ \textit{in vivo}
intracellular recorded neurons in the somatosensory cortex of the
rat. 
The spontaneous 
activity of each of these neurons shows negative serial interval
correlations \cite{nawrot_serial_2007} where the empirical sum over
correlation coefficients amounts to an average
$\sum_{i=1}^{10}\xi_i=-0.28$. We construct the population activity by
superimposing all 5 spike trains.
Thereafter, we estimate the Fano factor 
as a function of the observation time and compare it
to the case where, prior to superposition, renewal statistics is
enforced for each individual neuron through interval shuffling. Our
experimental observation in Fig.~\ref{result} (Right) confirms the
theoretical prediction of a reduced Fano factor simular to individual neurons 
\cite{Ratnam_Nonrenewal_2000} in the population level.



\textit{Benefits for Neural Coding.}  We provide three arguments that
demonstrate how the mechanism of spike-frequency adaptation benefits
neural processing and population coding.
First, our result of a reduced Fano factor $\check{F}<C_v^2$ for the
population activity of stationary adaptive processes ($bq>0$) directly
implies a reduction of the noise in the neuronal population rate code.
%
Our analysis of a set of cortical data suggests a reduction of $>50\%$
for long observation times.
The reduction of $J_T$ in Fig.~\ref{result}
becomes significant even for small observation times of $\approx 2$
average intervals, which 
is a relevant time scale for the transmission of a population rate
signal.
%
%
This result is reminiscent of an effect that has previously been
acknowledged as noise shaping and weak stimuli detection expressed in
the reduction of the low frequency power in a spectral analysis of
spike trains with negative serial interval correlations
\cite{chacron_noise_2004,*fuwape_spontaneous_2008,*chacron_negative_2001}. Our
result confirms their findings at the population level.


Our second argument is concerned with the transmission of a population
rate signal.
We may define a functional neural ensemble by the common postsynaptic
target neuron that receives the converging input of all
ensemble members. To elucidate the postsynaptic effect of adaptation
we simplify the ensemble autocorrelation function $A(u)$
following \cite{moreno-bote_theory_2008} with an exponential approximation
%
\begin{equation}
\textstyle \hat{A}(u) = r_{eq}\delta(u) + [ (F -1)/ 2\tau_c]
\exp(-u/\tau_c),
\label{A_approx}\end{equation}
where the second term is the approximation of $r_{eq}\mathcal{A}(u)$.
For given observation time window
$u$,
and $\tau_c$ the reduction of $F$ implies that $\hat{A}^{r}_u <
\hat{A}_u $. 
 Therefore, the
postsynaptic 
neuron receives inputs from an adaptive
ensemble that expresses an extended autocorrelation structure as
compared to the inputs from a non-adaptive ensemble.
%
%
Following the theory on the effect of input autocorrelation on signal
transmission in spiking neurons as developed in
\cite{moreno-bote_theory_2008, *moreno_response_2002}, a longer $\tau_c$ reduces
the input current fluctuations and this
facilitates a faster and more reliable
transmission of the modulated input rate signal by the postsynaptic
target neuron.

Finally we argue that a postsynaptic neuron can better
decode a small change in its input 
if the presynaptic neurons are adaptive.
To this end, we compute the information gain of the postsynaptic
activity,
between two counting distributions  
of an adaptive presynaptic ensemble when $\hat{h}_x(x)$ is
adiabatically transferred to $\hat{h}_x(x-\epsilon)$ with a small
change $\epsilon$ in the input ensemble. 
It is convenient to use $\tilde{\rho}_n(s)$ which associated
with counting distribution $\tilde{P}(n, s)$.
Thus,
we apply the Kullback-Leibler divergence to the Eq.~\eqref{ingenious2}
before and after the adiabatic change in the input
%
%
%
\begin{equation}
\textstyle D_{KL}( \tilde{\rho}_n^{\epsilon} || \tilde{\rho}_n) =
\sum_i \tilde{\rho}^{\epsilon}_{i}(s) \ln(
\tilde{\rho}^{\epsilon}_{i}(s)/ \tilde{\rho}_{i}(s)).
\label{KL} 
\end{equation}
%
%
Using Eq.~\eqref{auto} we obtain $D_{KL}( \tilde{\rho}_n^{\epsilon} ||
\tilde{\rho}_n) = \mathcal{A}^\epsilon_s
      [\ln(\mathcal{A}^\epsilon_s/\mathcal{A}_s)] $.
Due to  Eqs.~\eqref{SFAdynamics} and~\eqref{r_con},
the mean adaptation after the change is $E[x^\epsilon] = \tau q
r^\epsilon_{eq}$. If $\epsilon >0$ it follows that $r^\epsilon_{eq}
\geq r_{eq}$. Therefore the mean adaptation level increases and the
adapted process exhibits stronger negative serial correlations and
$\mathcal{A}^\epsilon_s >\mathcal{A}_s$. Thus, by
Eq.~\eqref{A_approx}, it is straight forward to deduce that
$D_{KL}>D^{r}_{KL}$, for renewal and adaptive processes with
identical interval distributions.

We now compute the information gain of the adaptive ensemble process
relative to a matched
Poisson rate model.
For different initial rate values $r_{eq}$ we assume a small but fixed
increase $\epsilon$ in the input that we express in parameter changes
$a^{\epsilon}$ and $b^{\epsilon}$ in Eq.~\eqref{adap_h} as outlined in
\cite{[{Supplementary material at}] supp}. This leads to an increase
$\kappa=r^{\epsilon}_{eq}-r_{eq}$ in rate that is effectively constant
over a wide range of initial values $r_{eq}$ (Fig.~2, Left).
In the rate model, assuming the same initial value of $r_{eq}$, the
same input step leads to a higher equilibrium rate increase
$\kappa^{Poisson}>\kappa$, which depends on the inital rate (Fig.~2,
Left) because the rate model lacks a mechanism of self-inhibition,
which in the adaptive model counteracts the rate increase.
%
%
%
Thereafter, we compute the Kullback-Leibler divergence
for both models and normalize it by the change in the output rate
$\kappa$. The result in Fig.~2 (Right) shows that $D_{KL}/ \kappa$ 
%
is larger for the adaptive model than for the rate model
%
across the range of tested input rates.
%
%
%
%
%
%
%
%
Thus, the information per extra spike is larger in the adaptive
ensemble than in the renewal ensemble,
and a postsynaptic neuron can discriminate small changes $\epsilon$
more efficiently, even though the absolute change in firing rate is
lower.

\begin{figure}
\includegraphics[scale=1.0]{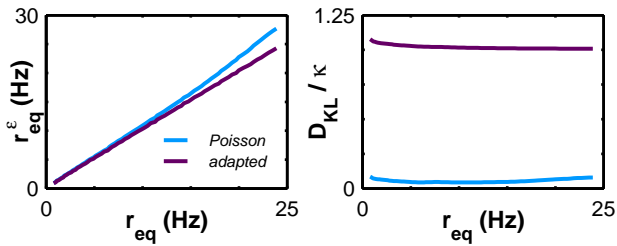}
\caption{Information gain per spike due to adaptation. \textbf{Left}:
  Transfer of equilibrium rate for fixed $\epsilon$ change of the
  input in adaptive and Poisson model. \textbf{Right}:
  Kullback-Leibler Divergence per extra spike as the measure of
  information gain for $n^\mathrm{th}$ event density of adaptive and
  Poisson processes while $u=200\mathrm{ms}$ and
  $\epsilon=0.01\mathrm{nS}$ with the same initial $r_{eq}$ and
  $\kappa = r_{eq}^{\epsilon}-r_{eq}$.
}
\label{result2}
\end{figure}

\textit{Discussion.} 
Our results point out a new aspect of spike frequency adaptation that
benefits
the reliable transmission and postsynaptic decoding of the neural
population code. 
This aspect adds to the known properties of compression and temporal
filtering of sensory input signals \cite{Lundstrom_fractional_2008} in
spike frequency adapting neurons.
%
The specific result of Eq.~\eqref{my_golden}
is also of practical consequence for the empirical analysis of the
so-called multi-unit activity.
By estimating Fano factor and serial correlations
we readily obtain an estimate of the average $C_v$ and serial correlation
of the individual processes.
%

 We developed a new formalism to treat event emitting systems that are
 influenced by a slow state variable, and we provide a number of
 useful general results on the higher order event statistics of
 superimposed renewal and non-renewal event processes, which are
 applicable to a wide range of event-based systems in
 nature~\cite{[{Supplementary material at}] supp}. The derivation of
 the state dependent hazard and master equation
 \cite{muller_spike-frequency_2007} assumes incoherent input
 fluctuation as in the mean-field theory, where common input is
 negligible. Treating a network with coherent fluctuations as
 encountered in finite size networks requires an alternative
 derivation of the hazard function~\cite{[{Supplementary material at}]
   supp}.

\textit{Acknowledgments}. We thank Carl van Vreeswijk and
 Stefano Cardanobile for valuable comments, and Clemens Boucsein
 for making the data available for re-analysis.
This work was funded by the German Ministry of Education and Research
to BCCN Berlin.


\bibliography{refs}

\begin{thebibliography}{25}%
\makeatletter
\providecommand \@ifxundefined [1]{%
 \@ifx{#1\undefined}
}%
\providecommand \@ifnum [1]{%
 \ifnum #1\expandafter \@firstoftwo
 \else \expandafter \@secondoftwo
 \fi
}%
\providecommand \@ifx [1]{%
 \ifx #1\expandafter \@firstoftwo
 \else \expandafter \@secondoftwo
 \fi
}%
\providecommand \natexlab [1]{#1}%
\providecommand \enquote  [1]{``#1''}%
\providecommand \bibnamefont  [1]{#1}%
\providecommand \bibfnamefont [1]{#1}%
\providecommand \citenamefont [1]{#1}%
\providecommand \href@noop [0]{\@secondoftwo}%
\providecommand \href [0]{\begingroup \@sanitize@url \@href}%
\providecommand \@href[1]{\@@startlink{#1}\@@href}%
\providecommand \@@href[1]{\endgroup#1\@@endlink}%
\providecommand \@sanitize@url [0]{\catcode `\\12\catcode `\$12\catcode
  `\&12\catcode `\#12\catcode `\^12\catcode `\_12\catcode `\%12\relax}%
\providecommand \@@startlink[1]{}%
\providecommand \@@endlink[0]{}%
\providecommand \url  [0]{\begingroup\@sanitize@url \@url }%
\providecommand \@url [1]{\endgroup\@href {#1}{\urlprefix }}%
\providecommand \urlprefix  [0]{URL }%
\providecommand \Eprint [0]{\href }%
\providecommand \doibase [0]{http://dx.doi.org/}%
\providecommand \selectlanguage [0]{\@gobble}%
\providecommand \bibinfo  [0]{\@secondoftwo}%
\providecommand \bibfield  [0]{\@secondoftwo}%
\providecommand \translation [1]{[#1]}%
\providecommand \BibitemOpen [0]{}%
\providecommand \bibitemStop [0]{}%
\providecommand \bibitemNoStop [0]{.\EOS\space}%
\providecommand \EOS [0]{\spacefactor3000\relax}%
\providecommand \BibitemShut  [1]{\csname bibitem#1\endcsname}%
\let\auto@bib@innerbib\@empty
\bibitem [{\citenamefont {Soler}\ \emph {et~al.}(2008)\citenamefont {Soler},
  \citenamefont {Rodriguez},\ and\ \citenamefont
  {Zumofen}}]{soler_memory_2008}%
  \BibitemOpen
  \bibfield  {author} {\bibinfo {author} {\bibfnamefont {F.~C.}\ \bibnamefont
  {Soler}}, \bibinfo {author} {\bibfnamefont {F.~J.}\ \bibnamefont
  {Rodriguez}}, \ and\ \bibinfo {author} {\bibfnamefont {G.}~\bibnamefont
  {Zumofen}},\ }\href {\doibase 10.1103/PhysRevA.78.053813} {\bibfield
  {journal} {\bibinfo  {journal} {Phys. Rev. A}\ }\textbf {\bibinfo {volume}
  {78}},\ \bibinfo {pages} {053813} (\bibinfo {year} {2008})}\BibitemShut
  {NoStop}%
\bibitem [{\citenamefont {Livina}\ \emph {et~al.}(2005)\citenamefont {Livina},
  \citenamefont {Havlin},\ and\ \citenamefont {Bunde}}]{livina_memory_2005}%
  \BibitemOpen
  \bibfield  {author} {\bibinfo {author} {\bibfnamefont {V.~N.}\ \bibnamefont
  {Livina}}, \bibinfo {author} {\bibfnamefont {S.}~\bibnamefont {Havlin}}, \
  and\ \bibinfo {author} {\bibfnamefont {A.}~\bibnamefont {Bunde}},\ }\href
  {\doibase 10.1103/PhysRevLett.95.208501} {\bibfield  {journal} {\bibinfo
  {journal} {Phys. Rev. Lett.}\ }\textbf {\bibinfo {volume} {95}},\ \bibinfo
  {pages} {208501} (\bibinfo {year} {2005})}\BibitemShut {NoStop}%
\bibitem [{\citenamefont {Lindner}\ and\ \citenamefont
  {Schwalger}(2007)}]{lindner_correlations_2007}%
  \BibitemOpen
  \bibfield  {author} {\bibinfo {author} {\bibfnamefont {B.}~\bibnamefont
  {Lindner}}\ and\ \bibinfo {author} {\bibfnamefont {T.}~\bibnamefont
  {Schwalger}},\ }\href@noop {} {\bibfield  {journal} {\bibinfo  {journal}
  {Phys. Rev. Lett.}\ }\textbf {\bibinfo {volume} {98}},\ \bibinfo {pages}
  {210603} (\bibinfo {year} {2007})}\BibitemShut {NoStop}%
\bibitem [{\citenamefont {Liu}\ and\ \citenamefont
  {Wang}(2001)}]{liu_spike-frequency_2001}%
  \BibitemOpen
  \bibfield  {author} {\bibinfo {author} {\bibfnamefont {Y.~H.}\ \bibnamefont
  {Liu}}\ and\ \bibinfo {author} {\bibfnamefont {X.~J.}\ \bibnamefont {Wang}},\
  }\href@noop {} {\bibfield  {journal} {\bibinfo  {journal} {J Comput
  Neurosci}\ }\textbf {\bibinfo {volume} {10}},\ \bibinfo {pages} {25}
  (\bibinfo {year} {2001})}\BibitemShut {NoStop}%
\bibitem [{\citenamefont {Farkhooi}\ \emph {et~al.}(2009)\citenamefont
  {Farkhooi}, \citenamefont {{Strube-Bloss}},\ and\ \citenamefont
  {Nawrot}}]{farkhooi_serial_2009}%
  \BibitemOpen
  \bibfield  {author} {\bibinfo {author} {\bibfnamefont {F.}~\bibnamefont
  {Farkhooi}}, \bibinfo {author} {\bibfnamefont {M.~F.}\ \bibnamefont
  {{Strube-Bloss}}}, \ and\ \bibinfo {author} {\bibfnamefont {M.~P.}\
  \bibnamefont {Nawrot}},\ }\href {\doibase 10.1103/PhysRevE.79.021905}
  {\bibfield  {journal} {\bibinfo  {journal} {Phys. Rev. E}\ }\textbf {\bibinfo
  {volume} {79}},\ \bibinfo {pages} {021905} (\bibinfo {year}
  {2009})}\BibitemShut {NoStop}%
\bibitem [{sup()}]{supp}%
  \BibitemOpen
  \href {http://link.aps.org/doi} {\ }\BibitemShut {NoStop}%
\bibitem [{\citenamefont {Muller}\ \emph {et~al.}(2007)\citenamefont {Muller},
  \citenamefont {Buesing}, \citenamefont {Schemmel},\ and\ \citenamefont
  {Meier}}]{muller_spike-frequency_2007}%
  \BibitemOpen
  \bibfield  {author} {\bibinfo {author} {\bibfnamefont {E.}~\bibnamefont
  {Muller}}, \bibinfo {author} {\bibfnamefont {L.}~\bibnamefont {Buesing}},
  \bibinfo {author} {\bibfnamefont {J.}~\bibnamefont {Schemmel}}, \ and\
  \bibinfo {author} {\bibfnamefont {K.}~\bibnamefont {Meier}},\ }\href
  {\doibase 10.1162/neco.2007.19.11.2958} {\bibfield  {journal} {\bibinfo
  {journal} {Neural Comp.}\ }\textbf {\bibinfo {volume} {19}},\ \bibinfo
  {pages} {2958} (\bibinfo {year} {2007})}\BibitemShut {NoStop}%
\bibitem [{\citenamefont {Gardiner}(2004)}]{gardiner_book}%
  \BibitemOpen
  \bibfield  {author} {\bibinfo {author} {\bibfnamefont {C.~W.}\ \bibnamefont
  {Gardiner}},\ }\href@noop {} {\emph {\bibinfo {title} {Handbook of Stochastic
  Methods}}},\ \bibinfo {edition} {3rd}\ ed.\ (\bibinfo  {publisher}
  {Springer},\ \bibinfo {year} {2004})\BibitemShut {NoStop}%
\bibitem [{\citenamefont {{de Kamps}}(2003)}]{dekamps_2003}%
  \BibitemOpen
  \bibfield  {author} {\bibinfo {author} {\bibfnamefont {M.}~\bibnamefont {{de
  Kamps}}},\ }\href {\doibase 10.1162/089976603322297322} {\bibfield  {journal}
  {\bibinfo  {journal} {Neural Comp.}\ }\textbf {\bibinfo {volume} {15}},\
  \bibinfo {pages} {2129} (\bibinfo {year} {2003})}\BibitemShut {NoStop}%
\bibitem [{\citenamefont {Gerstner}\ and\ \citenamefont
  {Kistler}(2002)}]{gerstner_spiking_2002}%
  \BibitemOpen
  \bibfield  {author} {\bibinfo {author} {\bibfnamefont {W.}~\bibnamefont
  {Gerstner}}\ and\ \bibinfo {author} {\bibfnamefont {W.~M.}\ \bibnamefont
  {Kistler}},\ }\href@noop {} {\emph {\bibinfo {title} {Spiking Neuron
  Models}}},\ \bibinfo {edition} {1st}\ ed.\ (\bibinfo  {publisher} {Cambridge
  University Press},\ \bibinfo {year} {2002})\BibitemShut {NoStop}%
\bibitem [{\citenamefont {{van
  Vreeswijk}}(2010)}]{van_vreeswijk_stochastic_2010}%
  \BibitemOpen
  \bibfield  {author} {\bibinfo {author} {\bibfnamefont {C.}~\bibnamefont {{van
  Vreeswijk}}},\ }in\ \href@noop {} {\emph {\bibinfo {booktitle} {Analysis of
  Parallel Spike Trains}}},\ \bibinfo {series and number} {Springer Series in
  Comput. Neurosci.}\ (\bibinfo  {publisher} {Springer},\ \bibinfo {year}
  {2010})\BibitemShut {NoStop}%
\bibitem [{\citenamefont {{McFadden}}(1962)}]{mcfadden_lengths_1962}%
  \BibitemOpen
  \bibfield  {author} {\bibinfo {author} {\bibfnamefont {J.~A.}\ \bibnamefont
  {{McFadden}}},\ }\href@noop {} {\bibfield  {journal} {\bibinfo  {journal} {J.
  of the Royal Stat. Soc. Series B}\ }\textbf {\bibinfo {volume} {24}},\
  \bibinfo {pages} {364} (\bibinfo {year} {1962})}\BibitemShut {NoStop}%
\bibitem [{\citenamefont {Cox}\ and\ \citenamefont
  {Lewis}(1966)}]{cox_statistical_1966}%
  \BibitemOpen
  \bibfield  {author} {\bibinfo {author} {\bibfnamefont {D.~R.}\ \bibnamefont
  {Cox}}\ and\ \bibinfo {author} {\bibfnamefont {P.~A.~W.}\ \bibnamefont
  {Lewis}},\ }\href@noop {} {\emph {\bibinfo {title} {The Statistical Analysis
  of Series of Events}}},\ Methuen's Monographs on Applied Probability and
  Statistics\ (\bibinfo  {publisher} {Methuen},\ \bibinfo {address} {London},\
  \bibinfo {year} {1966})\BibitemShut {NoStop}%
\bibitem [{\citenamefont {Cox}\ and\ \citenamefont
  {Isham}(1980)}]{cox_point_1980}%
  \BibitemOpen
  \bibfield  {author} {\bibinfo {author} {\bibfnamefont {D.}~\bibnamefont
  {Cox}}\ and\ \bibinfo {author} {\bibfnamefont {V.}~\bibnamefont {Isham}},\
  }\href@noop {} {\emph {\bibinfo {title} {Point Processes}}},\ \bibinfo
  {edition} {1st}\ ed.\ (\bibinfo  {publisher} {Chapman and {Hall/CRC}},\
  \bibinfo {year} {1980})\BibitemShut {NoStop}%
\bibitem [{\citenamefont {Lawrence}(1973)}]{lawrence_dependency_1973}%
  \BibitemOpen
  \bibfield  {author} {\bibinfo {author} {\bibfnamefont {A.~J.}\ \bibnamefont
  {Lawrence}},\ }\href@noop {} {\bibfield  {journal} {\bibinfo  {journal} {J.
  of the Royal Stat. Soc. Series B}\ }\textbf {\bibinfo {volume} {35}},\
  \bibinfo {pages} {306} (\bibinfo {year} {1973})}\BibitemShut {NoStop}%
\bibitem [{\citenamefont {C{\^a}teau}\ and\ \citenamefont
  {Reyes}(2006)}]{cteau_relation_2006}%
  \BibitemOpen
  \bibfield  {author} {\bibinfo {author} {\bibfnamefont {H.}~\bibnamefont
  {C{\^a}teau}}\ and\ \bibinfo {author} {\bibfnamefont {A.~D.}\ \bibnamefont
  {Reyes}},\ }\href {\doibase 10.1103/PhysRevLett.96.058101} {\bibfield
  {journal} {\bibinfo  {journal} {Phys. Rev. Lett.}\ }\textbf {\bibinfo
  {volume} {96}},\ \bibinfo {pages} {058101} (\bibinfo {year}
  {2006})}\BibitemShut {NoStop}%
\bibitem [{\citenamefont {Lindner}(2006)}]{lindner_superposition_2006}%
  \BibitemOpen
  \bibfield  {author} {\bibinfo {author} {\bibfnamefont {B.}~\bibnamefont
  {Lindner}},\ }\href {\doibase 10.1103/PhysRevE.73.022901} {\bibfield
  {journal} {\bibinfo  {journal} {Phys. Rev. E}\ }\textbf {\bibinfo {volume}
  {73}},\ \bibinfo {pages} {022901} (\bibinfo {year} {2006})}\BibitemShut
  {NoStop}%
\bibitem [{\citenamefont {Nawrot}\ \emph {et~al.}(2007)\citenamefont {Nawrot},
  \citenamefont {Boucsein}, \citenamefont {{Rodriguez-Molina}}, \citenamefont
  {Aertsen}, \citenamefont {Gruen},\ and\ \citenamefont
  {Rotter}}]{nawrot_serial_2007}%
  \BibitemOpen
  \bibfield  {author} {\bibinfo {author} {\bibfnamefont {M.~P.}\ \bibnamefont
  {Nawrot}}, \bibinfo {author} {\bibfnamefont {C.}~\bibnamefont {Boucsein}},
  \bibinfo {author} {\bibfnamefont {V.}~\bibnamefont {{Rodriguez-Molina}}},
  \bibinfo {author} {\bibfnamefont {A.}~\bibnamefont {Aertsen}}, \bibinfo
  {author} {\bibfnamefont {S.}~\bibnamefont {Gruen}}, \ and\ \bibinfo {author}
  {\bibfnamefont {S.}~\bibnamefont {Rotter}},\ }\href {\doibase
  10.1016/j.neucom.2006.10.101} {\bibfield  {journal} {\bibinfo  {journal}
  {Neurocomputing}\ }\textbf {\bibinfo {volume} {70}},\ \bibinfo {pages} {1717}
  (\bibinfo {year} {2007})}\BibitemShut {NoStop}%
\bibitem [{\citenamefont {Ratnam}\ and\ \citenamefont
  {Nelson}(2000)}]{Ratnam_Nonrenewal_2000}%
  \BibitemOpen
  \bibfield  {author} {\bibinfo {author} {\bibfnamefont {R.}~\bibnamefont
  {Ratnam}}\ and\ \bibinfo {author} {\bibfnamefont {M.~E.}\ \bibnamefont
  {Nelson}},\ }\href {http://www.jneurosci.org/cgi/content/abstract/20/17/6672}
  {\bibfield  {journal} {\bibinfo  {journal} {J. Neurosci.}\ }\textbf {\bibinfo
  {volume} {20}},\ \bibinfo {pages} {6672} (\bibinfo {year}
  {2000})}\BibitemShut {NoStop}%
\bibitem [{\citenamefont {Chacron}\ \emph {et~al.}(2004)\citenamefont
  {Chacron}, \citenamefont {Lindner},\ and\ \citenamefont
  {Longtin}}]{chacron_noise_2004}%
  \BibitemOpen
  \bibfield  {author} {\bibinfo {author} {\bibfnamefont {M.~J.}\ \bibnamefont
  {Chacron}}, \bibinfo {author} {\bibfnamefont {B.}~\bibnamefont {Lindner}}, \
  and\ \bibinfo {author} {\bibfnamefont {A.}~\bibnamefont {Longtin}},\
  }\href@noop {} {\bibfield  {journal} {\bibinfo  {journal} {Phys. Rev. Lett.}\
  }\textbf {\bibinfo {volume} {92}},\ \bibinfo {pages} {080601} (\bibinfo
  {year} {2004})}\BibitemShut {NoStop}%
\bibitem [{\citenamefont {Fuwape}\ and\ \citenamefont
  {Neiman}(2008)}]{fuwape_spontaneous_2008}%
  \BibitemOpen
  \bibfield  {author} {\bibinfo {author} {\bibfnamefont {I.}~\bibnamefont
  {Fuwape}}\ and\ \bibinfo {author} {\bibfnamefont {A.~B.}\ \bibnamefont
  {Neiman}},\ }\href {\doibase 10.1103/PhysRevE.78.051922} {\bibfield
  {journal} {\bibinfo  {journal} {Phys. Rev. E}\ }\textbf {\bibinfo {volume}
  {78}},\ \bibinfo {pages} {051922} (\bibinfo {year} {2008})}\BibitemShut
  {NoStop}%
\bibitem [{\citenamefont {Chacron}\ \emph {et~al.}(2001)\citenamefont
  {Chacron}, \citenamefont {Longtin},\ and\ \citenamefont
  {Maler}}]{chacron_negative_2001}%
  \BibitemOpen
  \bibfield  {author} {\bibinfo {author} {\bibfnamefont {M.~J.}\ \bibnamefont
  {Chacron}}, \bibinfo {author} {\bibfnamefont {A.}~\bibnamefont {Longtin}}, \
  and\ \bibinfo {author} {\bibfnamefont {L.}~\bibnamefont {Maler}},\
  }\href@noop {} {\bibfield  {journal} {\bibinfo  {journal} {J. Neurosci.}\
  }\textbf {\bibinfo {volume} {21(14)}},\ \bibinfo {pages} {5328} (\bibinfo
  {year} {2001})}\BibitemShut {NoStop}%
\bibitem [{\citenamefont {{Moreno-Bote}}\ \emph {et~al.}(2008)\citenamefont
  {{Moreno-Bote}}, \citenamefont {Renart},\ and\ \citenamefont
  {Parga}}]{moreno-bote_theory_2008}%
  \BibitemOpen
  \bibfield  {author} {\bibinfo {author} {\bibfnamefont {R.}~\bibnamefont
  {{Moreno-Bote}}}, \bibinfo {author} {\bibfnamefont {A.}~\bibnamefont
  {Renart}}, \ and\ \bibinfo {author} {\bibfnamefont {N.}~\bibnamefont
  {Parga}},\ }\href {\doibase 10.1162/neco.2008.03-07-497} {\bibfield
  {journal} {\bibinfo  {journal} {Neural Comput}\ }\textbf {\bibinfo {volume}
  {20}},\ \bibinfo {pages} {1651} (\bibinfo {year} {2008})}\BibitemShut
  {NoStop}%
\bibitem [{\citenamefont {Moreno}\ \emph {et~al.}(2002)\citenamefont {Moreno},
  \citenamefont {de~la Rocha}, \citenamefont {Renart},\ and\ \citenamefont
  {Parga}}]{moreno_response_2002}%
  \BibitemOpen
  \bibfield  {author} {\bibinfo {author} {\bibfnamefont {R.}~\bibnamefont
  {Moreno}}, \bibinfo {author} {\bibfnamefont {J.}~\bibnamefont {de~la Rocha}},
  \bibinfo {author} {\bibfnamefont {A.}~\bibnamefont {Renart}}, \ and\ \bibinfo
  {author} {\bibfnamefont {N.}~\bibnamefont {Parga}},\ }\href {\doibase
  10.1103/PhysRevLett.89.288101} {\bibfield  {journal} {\bibinfo  {journal}
  {Phys. Rev. Lett.}\ }\textbf {\bibinfo {volume} {89}},\ \bibinfo {pages}
  {288101} (\bibinfo {year} {2002})}\BibitemShut {NoStop}%
\bibitem [{\citenamefont {Lundstrom}\ \emph {et~al.}(2008)\citenamefont
  {Lundstrom}, \citenamefont {Higgs}, \citenamefont {Spain},\ and\
  \citenamefont {Fairhall}}]{Lundstrom_fractional_2008}%
  \BibitemOpen
  \bibfield  {author} {\bibinfo {author} {\bibfnamefont {B.~N.}\ \bibnamefont
  {Lundstrom}}, \bibinfo {author} {\bibfnamefont {M.~H.}\ \bibnamefont
  {Higgs}}, \bibinfo {author} {\bibfnamefont {W.~J.}\ \bibnamefont {Spain}}, \
  and\ \bibinfo {author} {\bibfnamefont {A.~L.}\ \bibnamefont {Fairhall}},\
  }\href {\doibase 10.1038/nn.2212} {\bibfield  {journal} {\bibinfo  {journal}
  {Nat Neurosci}\ }\textbf {\bibinfo {volume} {11}},\ \bibinfo {pages} {1335}
  (\bibinfo {year} {2008})}\BibitemShut {NoStop}%
\end{thebibliography}%
\end{document}